\documentclass[10pt, conference, compsocconf]{IEEEtran}
\usepackage{color, colortbl}
\usepackage[table]{xcolor} 
\usepackage{multirow}
\usepackage{array, boldline, makecell, booktabs, ragged2e}
\usepackage{multicol}
\usepackage{boldline, makecell}
\usepackage{tfrupee}
\newcolumntype{P}[1]{>{\centering\arraybackslash}p{#1}}
\usepackage{color}
\definecolor{lgrey}{RGB}{140,140,140}
\definecolor{black}{RGB}{255, 255, 255}
\definecolor{vred}{RGB}{216, 191, 216}
\definecolor{skyblue}{RGB}{0, 27, 255}
\definecolor{tomato}{RGB}{255, 67, 0}
\usepackage{devanagari}
\usepackage{blindtext}
\usepackage{accsupp}
\usepackage{fontspec}
\usepackage{pifont}
\usepackage{svg}
\newfontscript{Devanagari}{deva,dev2}
\newfontface{\hindi}[Script=Devanagari]{Lohit-Devanagari.ttf}
\usepackage[noadjust]{cite}

\begin{document}
%
% paper title
% can use linebreaks \\ within to get better formatting as desired
\title{\textit{Vyaktitv}: A Multimodal Peer-to-Peer Hindi Conversations based Dataset for Personality Assessment}
% author names and affiliations
% use a multiple column layout for up to two different
% affiliations
% \author{\IEEEauthorblockN{Shahid Nawaz Khan, Maitree Leekha, Jainendra Shukla, Rajiv Ratn Shah}
% \IEEEauthorblockA{IIIT-Delhi, New Delhi, India}}

\author{\IEEEauthorblockN{Shahid Nawaz Khan\IEEEauthorrefmark{1},
Maitree Leekha\IEEEauthorrefmark{2}, Jainendra Shukla\IEEEauthorrefmark{1},
Rajiv Ratn Shah\IEEEauthorrefmark{1}}
\IEEEauthorblockA{\IEEEauthorrefmark{1}IIIT-Delhi, New Delhi, India}
\IEEEauthorblockA{\IEEEauthorrefmark{2}Delhi Technological University, New Delhi, India}
\IEEEauthorblockN{Email: \texttt{\small shahid17102@iiitd.ac.in, maitreeleekha\_bt2k16@dtu.ac.in,}} \IEEEauthorblockN{\texttt{\small jainendra@iiitd.ac.in, rajivratn@iiitd.ac.in}}}

\maketitle
\begin{abstract}
Automatically detecting personality traits can aid several applications, such as mental health recognition and human resource management. Most datasets introduced for personality detection so far have analyzed these traits for each individual in isolation. However, personality is intimately linked to our social behavior. Furthermore, surprisingly little research has focused on personality analysis using low resource languages. To this end, we present a novel peer-to-peer Hindi conversation dataset,  \textit{Vyaktitv}\footnote{In Hindi, \textit{Vyaktitv} ({\hindi व्यक्तित्व }) means \textit{personality}.}. It consists of high-quality audio and video recordings of the participants, with Hinglish\footnote{Hindi words are written using English alphabets.} textual transcriptions for each conversation. The dataset also contains a rich set of socio-demographic features, like income, cultural orientation, amongst several others, for all the participants. We release the dataset for public use, as well as perform preliminary statistical analysis along the different dimensions. Finally, we also discuss various other applications and tasks for which the dataset can be employed.   
\end{abstract}


\IEEEpeerreviewmaketitle

\section{Introduction}

According to psychologists, the study of the personality involves addressing some of the most exciting queries on human psychology, including analysis of how different aspects of people's lives are related, and how they interact with the environment as social beings \cite{persTheory}. All of it is an attempt towards better understanding and embracing the differences and uniqueness amongst individuals.

Several formal definitions of personality exist in the literature \cite{corr2009cambridge}. However, loosely speaking, an individual's personality primarily constitutes their social and emotional behavior, cognition, and decision-making abilities \cite{persTheory}. Analyzing these personal characteristics can help improve our understanding and management of people in a way that would not have been possible otherwise. Applications of personality assessment can be dated back to $1920$s, when it was used to supplement the process of personnel recruitment, particularly for armed forces, as well as to identify post-traumatic stress disorder (PTSD) symptoms \cite{mehta2019recent}. Even today, among several others, the National Aeronautics and Space Administration (NASA) uses different kinds of group activities to assess traits that are fundamental for survival in critical situations \cite{Mana2007MultimodalCO}. Therefore, automatic personality assessment can pave the way for substantial improvements in the aforementioned domains, among several others.

Several datasets have been introduced recently, each adopting a unique method for assessing personality, including the analysis of social media data \cite{Wang2015UnderstandingPT}, anonymous essays \cite{sun2018personality}, short video-based self-introductions \cite{yu2019speaking}, call-center simulations \cite{ivanov2011recognition}, performance in group discussions \cite{pianesi10.1145/1452392.1452404}, etc. While the studies using these datasets present exciting insights into their subjects' personalities, they have the following limitations.

\begin{itemize}
    % \item
    % Studies show that low-level multimodal signals perform better for assessing psychological characteristics \cite{multi-psycho}, yet many of them use unimodal visual \cite{berkovsky2019detecting, gavrilescu2015study}, vocal prosodic \cite{yu2019speaking} or textual \cite{Wang2015UnderstandingPT, sun2018personality} features only. %, with only a very few attempts at combining these three media forms for personality attempts.
    
    % CHANGE: CITE DATASETS INSTEAD OF STUDIES
    \item Firstly, a large proportion of them \cite{berkovsky2019detecting, gavrilescu2015study, yu2019speaking, Wang2015UnderstandingPT, sun2018personality, ivanov2011recognition, AHMAD20171964} miss an essential ingredient for personality assessment: a social environment. Our idiosyncrasies largely stem from the way we routinely interact with those around us \cite{dreier2011personality}.
    
    \item Another critical aspect is that a large body of work \cite{majumder2017deep, pianesi10.1145/1452392.1452404, Wang2015UnderstandingPT} has been devoted to personality assessment where the primary language used was English, with very little literature about the use of other languages. In other words, the advances made so far have not been tested for non-English speakers.
\end{itemize}

 To this end, we propose a novel peer-to-peer interaction-based multimodal dataset for assessment of personality traits. Our method differs from the previously used techniques of group discussions \cite{pianesi10.1145/1452392.1452404}, and simulation activities \cite{ivanov2011recognition} by analyzing individual behavior in an unrestrained setting, where the participants are at liberty to guide their conversations, react openly, and without any formal criteria of judgment that may exist in the former. Furthermore, we assess conversations where the participants interact in Hindi, which is used by over $637$ million \footnote{Hindi usage statistics from https://www.statista.com/statistics/266808/the-most-spoken-languages-worldwide/} people around the world. Finally, previous research \cite{goldberg1998demographic} has shown that factors like age, ethnicity, and educational qualification can significantly impact an individual's personality. We extend these findings by conducting an ethnographic study analyzing several socio-demographic indicators and how they impact an individual's personality. Our contributions are summarized as follows.

\begin{itemize}
    \item First multimodal peer-to-peer Hindi conversation-based personality assessment dataset. It includes data in three different media forms: $(1)$ video feed of the participants, $(2)$ their audio recordings,  and $(3)$ Hinglish transcriptions of the dialogue between the participants. 

    % \item A multimodal analysis of personality traits using visual, prosodic, and textual cues from the conversations. Our experiments using different combinations of these features reveal that multimodality boosts the performance of personality assessment.  
    \item Rich socio-demographic indicators for the participants involved in the study. Many of these, like social media usage, public speaking skills, cultural inclination, have not been considered in the past. Our statistical analysis reveals that these factors significantly impact a person's personality. 
\end{itemize}

% CHANGE THE ORGANIZATION
The rest of this paper has been organized as follows. The next section reviews some of the widely used theoretical models of personality and the techniques used for personality trait detection. Section~\ref{dataset} describes the data collection process, while Section~\ref{data_analysis}, continues with the statistical analysis of socio-demographic and lexical features, in terms of their relation with different personality traits. Finally, with Section~\ref{recc_fw}, we conclude our work and discuss some of the other applications of the rich, multimodal dataset proposed in this work.

\section{Literature Review}
\subsection{Personality Measures}
There are many theories in modern psychology modeling the personality traits of humans based on several interesting measures. We will be discussing some of them briefly. The interested reader may refer to psychological surveys \cite{doi:10.1177/0098628311411785} and books \cite{persTheory} for further details on the subject.

One of the oldest ways of modeling personality is the Trait Theory \cite{pervin1999handbook}, which aims at analyzing different human behaviors by treating them as `traits’, like \textit{Extraversion}, \textit{Emotional Stability}, etc. It usually follows a lexical approach, wherein the subjects use short phrases to discuss their behavior, based on which a score is given for different traits. More recently, Cattell \textit{et al.} \cite{cattell2008sixteen} proposed the 16PF model, where they used 16 personality factors, including \textit{Dominance}, \textit{Perfectionism}, \textit{Privateness}, and several others, to describe human behavior. 
Eysenck \cite{eysenck2012model} produced a more concise three-dimensional model for personality assessment, using \textit{Psychoticism}, \textit{Extraversion}, and \textit{Neuroticism} (PEN) as factors. Furthermore, questionnaires are often used to analyze personality traits based on the subjects’ responses to a set of specific questions. The Myers Briggs Type Indicator (MBTI) \cite{briggs2000introduction} is one such test widely used around the world. It is a self-introspective report which reflects on an individual’s decision-making process. Specifically, the test classifies the subjects into two categories, along four different dimensions, namely, \textit{Introversion}/\textit{Extraversion}, \textit{Sensing}/\textit{Intuition}, \textit{Thinking}/\textit{Feeling}, \textit{Judging}/\textit{Perception}. Finally, one of the most popular and widely accepted models in the literature, and the one we use as a foundation in our work, is the Five-Factor model, also known as the Big Five model \cite{digman1990personality}. As in the case of the Trait Theory, which analyses personality differences based on individual behaviors, the Big Five model is based on five personality factors of subjects, summarized in Table~\ref{bigfive_tab}.

\begin{table}[!h]
    \centering
    \resizebox{0.9\columnwidth}{!}{
    \begin{tabular}{c|c|c}
    \Xhline{1pt}
        \textbf{Personality Trait} & \textbf{High Score} & \textbf{Low Score} \\\hline
        \textbf{Extraversion} (\texttt{EXT}) & Extrovert & Introvert\\
        \textbf{Neuroticism} (\texttt{NEU}) & Anxious & Confident\\
        \textbf{Agreeableness} (\texttt{AGR}) & Trustworthy & Selfish\\
        \textbf{Conscientiousness} (\texttt{CON}) & Disciplined & Confused\\
         \textbf{Openness} (\texttt{OPN}) & Innovative & Unimaginative\\
     \Xhline{1pt}
    \end{tabular}}
    \vspace{2mm}
    \caption{Big Five personality traits, and the interpretations with high and low scores for these traits.}
    \label{bigfive_tab}
\end{table}

\subsection{Datasets for Personality Assessment}
There have been several studies analyzing human personality traits, and they differ substantially in the way they model the personality of individual subjects. Ranging from analyzing telephonic conversations of the participants \cite{ivanov2011recognition} to assessing these traits based on their resume \cite{cole2003using}, the experimental designs of these studies vary significantly. Therefore, this section focuses on datasets and experimental methods explored by the researchers to assess personality, and how our \textit{Vyaktitv} dataset differs from them.    

\vspace{2mm}
\noindent\textbf{Text based analysis.} In his study, Wang \cite{Wang2015UnderstandingPT} used social media data for predicting personality by analyzing the relationship between the language used by the users on these platforms and their personality traits. He curated a dataset of around 90,000 users and used their recent posts on Twitter to predict the MBTI \cite{briggs2000introduction} personality traits. Many researchers \cite{majumder2017deep, sun2018personality} have also analyzed personality traits based on essays written by the subjects. In this regard, a particularly popular benchmark dataset is the one by Pennebaker and King \cite{pennebaker1999linguistic}, used to predict the Big Five traits based on user-written essays. Automatic prediction of personality impressions based on the verbal content of vlogs has also attracted the research community's active interest \cite{biel2013hi}. 

\vspace{2mm}
\noindent\textbf{Audio and Vision based analysis.}
Ivanov \textit{et al.} \cite{ivanov2011recognition} curated the PersIA dataset, containing a series of telephonic conversations between participants simulating a call center. Furthermore, they also found the participants' verbal backchanneling responses to be useful predictors of their personality. Using vlogs for audio and video-based automatic assessment of personality has also been actively pursued. In this regard, the ChaLearn First Impressions dataset \cite{escalante2018explaining} has been widely experimented with. 

Early attempts focused on the use of facial expressions to predict personality. The Cohn-Kanade Dataset \cite{cohnemodataset} is a widely used repository for facial expressions, and has been used jointly with other personality-based datasets \cite{gavrilescu2015study} to predict the traits. In another study, Berkovsky \textit{et al.} \cite{berkovsky2019detecting} followed a different approach by analyzing personality traits based on the subjects' physiological responses to external stimuli. Specifically, they employed affective videos and images to evoke emotional responses in the participants and used eye-tracking data captured using Eye-Tracking Glasses (ETG) to predict the personality traits. Furthermore, there have also been attempts at predicting the personality traits in crowds. In particular,  Favaretto \textit{et al.} \cite{favaretto2019detecting} developed a video analysis system that could detect the personality, emotion, and cultural aspects of pedestrians using video sequences. 

\vspace{2mm}
\noindent\textbf{Use of Group Discussions.}
Several researchers have used group discussions to assess leadership qualities and personality traits. These include the MS-2 \cite{Mana2007MultimodalCO}, the MATRICS \cite{Nihei2014PredictingIS}, and the ELEA-AV \cite{eleaav2011nonverbal} corpuses. However, such setups restrain the free flow of ideas, where the discussion is likely to be guided by the leader, and the introverts may get less chance to speak.  Moreover, in discussions like these, people want their views to be accepted by the group, and emphasis is on winning \cite{senge1990fifth} (eg: being elected as the leader).\\ 

A very few studies, like the one conducted by Kalimeri \textit{et al.} \cite{wild},  have attempted to model the personality of individuals based on how they interact with their environment in their daily lives. In their experiment, Kalimeri \textit{et al.} used sensors to track the participants' activity over a period of six weeks. In the present study, we follow a different approach to mimic the social setting by making the subjects participate in peer-to-peer conversations and discuss topics from their daily lives. Peer-to-peer conversations have been utilized in a lot of different domains \cite{confpaperMMM, empatheticagents}; however, to the best of our knowledge, personality assessment research has not focused on such interactions. Furthermore, in most of the datasets included as a part of the literature, the primary language used was English. Our dataset, \textit{Vyaktitv}, stands as the first Hindi peer-to-peer conversation dataset and sets us apart from the previous studies. Finally, unlike most of these studies, we also collect the socio-demographic data of the participants involved in the study. Factors like gender and age have been known to impact personality traits \cite{goldberg1998demographic}. We gather several other features (like if subjects have a sibling or not, or if they live with their parents, etc.) and statistically analyze their influence on the Big Five traits.

\section{\textit{Vyaktitv} Dataset}
\label{dataset}

% -video
% -audio
% -transcriptions
% -socio demographic features
The \textit{Vyaktitv} dataset\footnote{Interested researchers can reach out to any of the authors to gain access to the \textit{Vyaktitv} dataset.} provides rich multimodal information from peer-to-peer interactions between individuals. The next few subsections discuss the data collection process, its main features, and the annotations done. %Specifically, it includes data of the following major categories, collected for both the participants involved in a conversation:
\subsection{Data Collection Methodology}
\noindent\textbf{Participants.}
A total of $38$ people ($24$ Male, $14$ Female) participated in the experiment. Their average age was $21.47$ years, with a standard deviation of $2.25$ years. Most of them were undergraduate or PhD students at an Indian university, while some others were employed as researchers and engineers. All the participants were native Hindi speakers, and had formally learnt Hindi through courses at their high school. 

\vspace{1mm}
\noindent\textbf{Procedure.} 
The participants were called randomly in pairs to interact with each other. Note, that a participant could be friends with some, and a total stranger to others. A random pairing process discarded the possibility of only friendly pairs being recruited for conversations. They were provided with some initial prompts for starting the conversation, as shown in Table~\ref{conv_prompts}, although they were also encouraged to choose a topic of their liking. Furthermore, they were encouraged to speak in Hindi at all times, but the mode of communication was a mix of Hindi and some bit of English. A total of $25$ conversations were recorded, where the average length of each was $16$ minutes and $6$ seconds, totaling to about $702$ minutes of multimodal content.

\begin{table}[!t]
    \centering
    \resizebox{\columnwidth}{!}{
    \begin{tabular}{cp{\columnwidth}}
    \Xhline{1pt}
    \rowcolor{lgrey!25}
    \textbf{\texttt{[P1]}} & In context of the Indian cricket, is Virat Kohli a better test-match captain than Mahendra Singh Dhoni?\\

   \textbf{\texttt{[P2]}} &  What are your favourite places to visit in India?\\
    
    \rowcolor{lgrey!25}
   \textbf{\texttt{[P3]}} &  Do you think pets are an important part of a family?\\

    \textbf{\texttt{[P4]}} & What are your plans after graduation?\\
    
    \rowcolor{lgrey!25}
    \textbf{\texttt{[P1]}} & In your opinion, should cell phones be allowed during class?\\

    \textbf{\texttt{[P5]}} & How do you think global warming is impacting our lives?\\
    
    \rowcolor{lgrey!25}
    \textbf{\texttt{[P6]}} & Should children be detained in high schools?\\
    
    \textbf{\texttt{[P7]}} & How do you feel about euthanasia (mercy killing)?\\
    
    \rowcolor{lgrey!25}
    \textbf{\texttt{[P8]}} & Do you think watching television can help in developing the minds of young children?\\

    \textbf{\texttt{[P9]}} & In your view, is cloning animals ethical?\\
   
   \rowcolor{lgrey!25}
    \textbf{\texttt{[P10]}} & Comment on the status of education in rural areas, and the high dropout rate among children of under-privileged families.\\
    
    \textbf{\texttt{[P11]}} & In your opinion, what are the most promising alternative sources of energy, especially for the Indian economy?\\
    \Xhline{1pt}
    \end{tabular}}
    \caption{Conversation prompts: Sample topics given to the participants to start a conversation. These prompts were introduced to the participants in Hindi as well.}
    \label{conv_prompts}
\end{table}

\begin{figure}[!t]
    \centering
    \includegraphics[width=0.8\columnwidth]{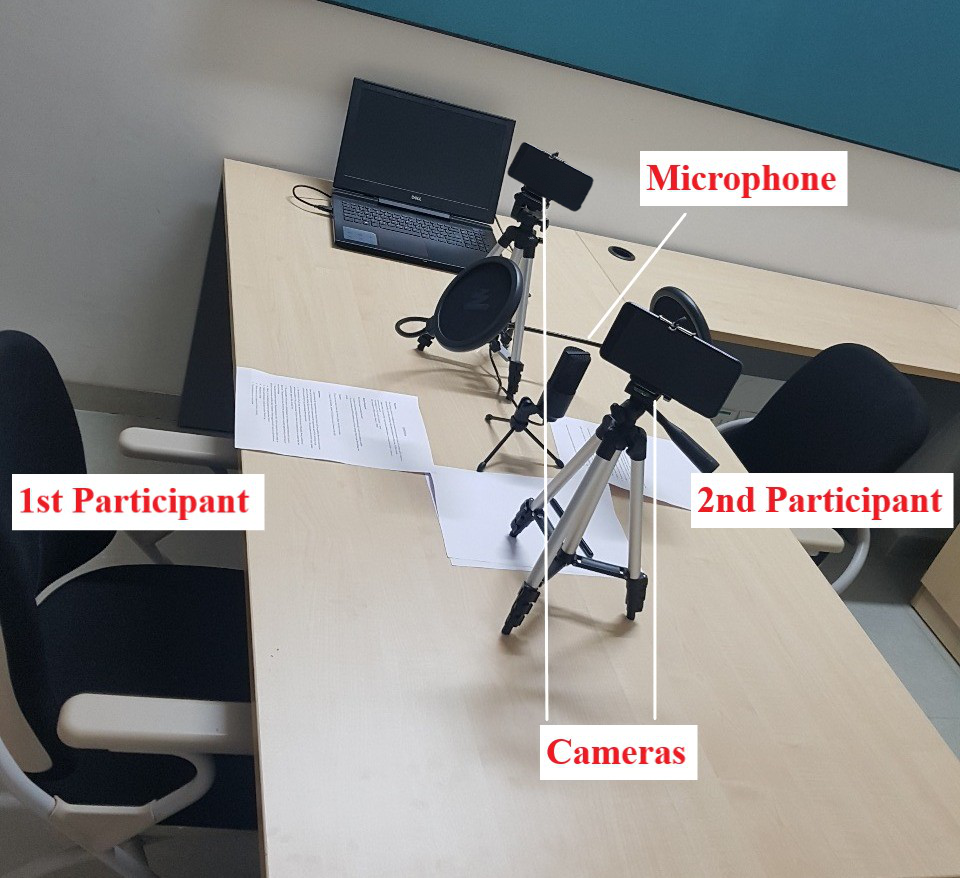}
    \caption{Experimental Setup: Two separate high quality cameras, and a microphone were used to record the video and audio of the two participants.}
    \label{expt_Setup}
    % \vspace{-5mm}
\end{figure}

\begin{table}[!t]
% \centering
% \vspace{2mm}
\resizebox{\columnwidth}{!}{
\begin{tabular}{p{\columnwidth}}
\Xhline{1pt}\\
\textbf{\texttt{[Q1]}} What best describes your gender? \vspace{0mm} \textcolor{lgrey}{(\texttt{Female}, \texttt{Male})}\vspace{2mm}\\
\textbf{\texttt{[Q2]}} How old are you? \vspace{0mm}\vspace{2mm}\\
\textbf{\texttt{[Q3]}} Please mention the amount of work experience you have had (in years). Enter 1 if the duration is < 1 year (say, as an intern). \vspace{2mm}\\
\textbf{\texttt{[Q4]}} What is your current state of employment? \\ \textcolor{lgrey}{(\texttt{Junior Undergraduate Student}, \texttt{Senior Undergraduate Student}, \texttt{M. Tech. Student}, \texttt{Junior PhD Student}, \texttt{Senior PhD Student}, \texttt{Employed}, \texttt{Unemployed})}\vspace{2mm}\\
\textbf{\texttt{[Q5]}} One a scale of 1 to 3, how would you rate your presence on social media, including platforms like Twitter, Facebook, and Instagram? By "presence", we mean how often do you post and reflect your thoughts on these platforms.\vspace{2mm}\\
\textbf{\texttt{[Q6]}} Do you have a sibling? 
\textcolor{lgrey}{(\texttt{Yes}, \texttt{No})} \vspace{0mm}\\
\textbf{\texttt{[Q7]}} Is your primary area of residence in the city (urban) or somewhere in the outskirts (rural)? \textcolor{lgrey}{(\texttt{Urban}, \texttt{Rural})} \vspace{2mm}\\
\textbf{\texttt{[Q8]}} Are you currently staying with your parents? \textcolor{lgrey}{(\texttt{Yes}, \texttt{No})} \vspace{2mm}\\
\textbf{\texttt{[Q9]}} What is your mother's highest qualification? \\ \textcolor{lgrey}{(\texttt{$<$ Class 10\textsuperscript{th}}, \texttt{Class 10\textsuperscript{th}}, \texttt{Class 12\textsuperscript{th}}, \texttt{Bachelor's/Diploma}, \texttt{Post Graduate/Professional Training})}\vspace{2mm}\\
\textbf{\texttt{[Q10]}} What is your father's highest qualification? \\ \textcolor{lgrey}{(\texttt{same options as for the previous question})} \vspace{2mm}\\
\textbf{\texttt{[Q11]}} When at home, do you use Hindi as the main language to converse? \textcolor{lgrey}{(\texttt{Yes}, \texttt{No})} \vspace{2mm}\\
\textbf{\texttt{[Q12]}} Do you think you are a religious/culturally inclined person? \textcolor{lgrey}{(\texttt{Yes}, \texttt{No})} \vspace{2mm}\\
\textbf{\texttt{[Q13]}} Please select the band that closes matches your family's annual income in INR (\rupee). \\ \textcolor{lgrey}{($<$ \rupee\texttt{0.5 million}, \rupee\texttt{0.5 million} - \rupee\texttt{1 million}, $>$ \rupee\texttt{1 million})}. In USD: \textcolor{lgrey}{($<$ \texttt{USD 6600}, \texttt{USD 6600} - \texttt{USD 13200}, $>$ \texttt{USD 13200}) } \vspace{2mm}\\
\textbf{\texttt{[Q14]}} On a scale of 1 to 5, how would rate your public speaking skills, say when formally addressing a group of people, or even with friends?  \vspace{2mm}\\
\textbf{\texttt{[Q15]}} On a scale of 1 to 5, how would rate your writing proficiency?\vspace{3mm}\\
\Xhline{1pt}

\end{tabular}}
\caption{Questionnaire for gathering socio-demographic features from the participants involved in the experiment.}
\label{socio-demo-ques} 
\end{table}

\subsection{Multimodal Content}
\textit{Vyaktitv} includes data of the following main categories, collected for both the participants involved in a conversation:

\vspace{2mm}
\noindent\textbf{Video and Audio Recordings.} We used two time-synchronized $48$ MegaPixel cameras to capture the front view of both the participants (Figure~\ref{expt_Setup}). A Fifine K669B Condenser Recording USB Microphone (with a sound-to-noise ratio of $78$ \texttt{dB}) was used to record the conversation. In addition, the dual mics of a Samsung Galaxy S7 Edge were used in interview mode to record each participant separately. Although the dual mics had active noise cancellation, additional processing was done using the open-source tool Audacity\footnote{https://www.audacityteam.org/} for further noise reduction. 
 
\vspace{1mm}
  \noindent\textbf{Hinglish Transcriptions.} We also provide manual transcriptions of the conversations in Hinglish. The transcribers were recruited from a university and were all native Hindi speakers. They had also received formal education for both Hindi and English at a high-school level.  \\
  \noindent \textit{Transcription Guidelines.} The audio files recorded using the dual mics were given to the transcribers with the following instructions:

\begin{itemize}
    \item Each file had speech signals from two speakers, one of which was more prominent than the other. The annotators were instructed to type the dialogues spoken only by the prominent speaker.
    
    \item The language to be used was Hinglish, \textit{i.e.}, Hindi words written using English alphabets.
    
    \item Previous research has established the utility of verbal backchannels for analyzing personality \cite{ivanov2011recognition}. Therefore, we manually annotated our Hinglish transcriptions for several such feedbacks. In particular, the transcribers were instructed to determine and annotate the lexical cues in shown in  Table~\ref{linguistic_annotations}. The annotators were introduced to these words and told to annotate any occurrence of such terms by appending ``xxx''. This made an automatic analysis of these words relatively easy. The transcribers were provided with a sample list of words from each category for reference.

    \end{itemize}

\begin{figure*}[!t]
\centering
\begin{minipage}{0.20\textwidth}
\centering
    \includegraphics[width=1.00\textwidth]{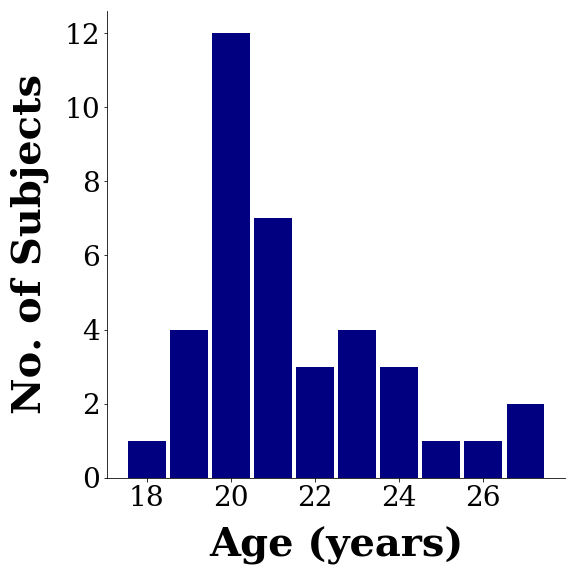}
    % \\$(i)$
\end{minipage}
\hfill
\begin{minipage}{0.16\textwidth}
\centering
    \includegraphics[width=1.0\textwidth]{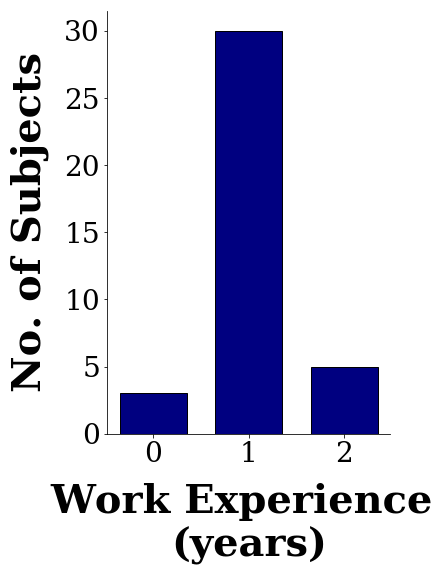}
    % \\$(ii)$
\end{minipage}
\hfill
\begin{minipage}{0.18\textwidth}
\centering
    \includegraphics[width=1.0\textwidth]{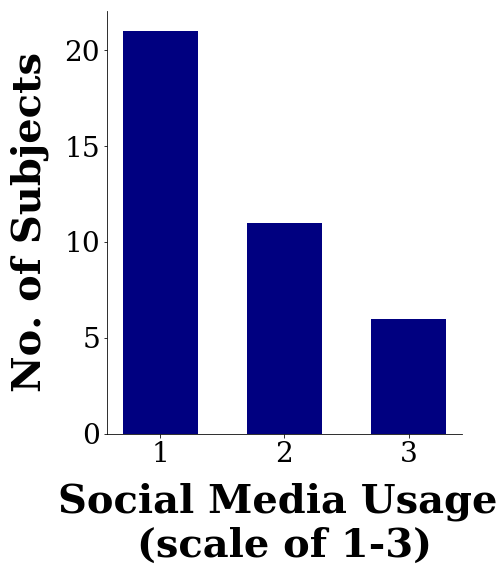}
    % \\$(iiii)$
\end{minipage}
\hfill
\begin{minipage}{0.13\textwidth}
\centering
    \includegraphics[width=1.0\textwidth]{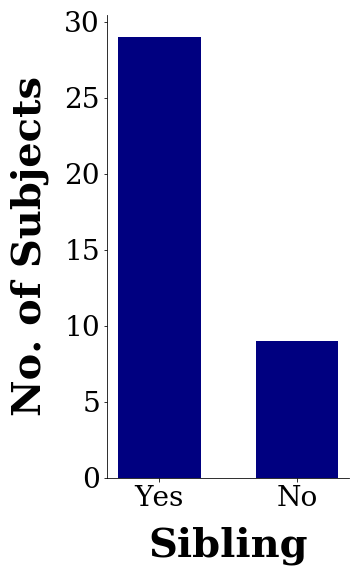}
    % \\$(iv)$
\end{minipage}
\hfill
\begin{minipage}{0.13\textwidth}
\centering
    \includegraphics[width=1.0\textwidth]{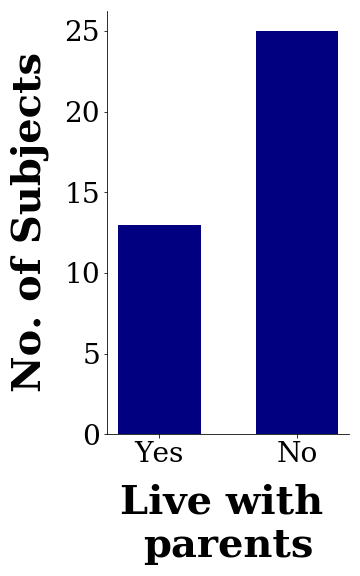}
    % \\$(v)$
\end{minipage}
\\
\hfill
\begin{minipage}{0.13\textwidth}
\centering
    \includegraphics[width=1.00\textwidth]{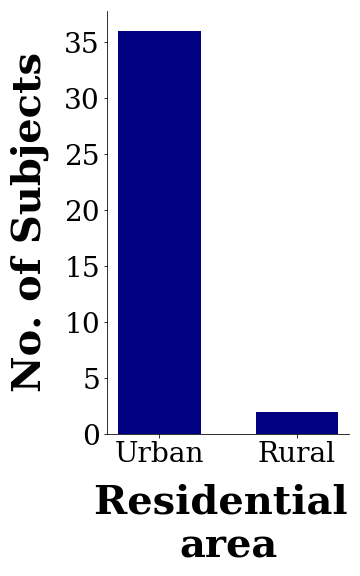}
    % \\$(vi)$
\end{minipage}
\hfill
\begin{minipage}{0.13\textwidth}
\centering
    \includegraphics[width=1.0\textwidth]{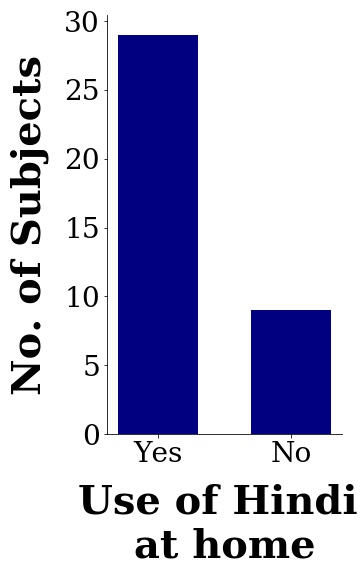}
    % \\$(vii)$
\end{minipage}
\hfill
\begin{minipage}{0.16\textwidth}
\centering
    \includegraphics[width=1.0\textwidth]{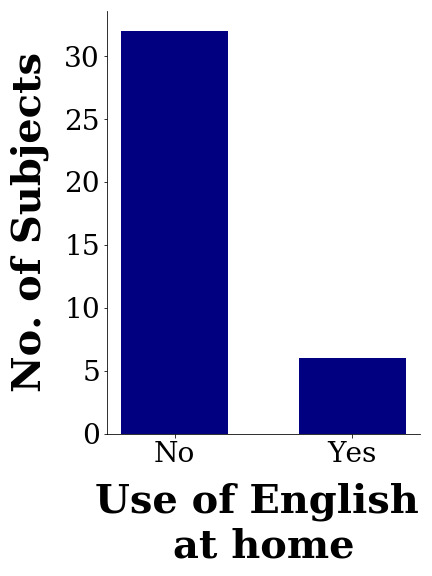}
    % \\$(viii)$
\end{minipage}
\hfill
\begin{minipage}{0.13\textwidth}
\centering
    \includegraphics[width=1.0\textwidth]{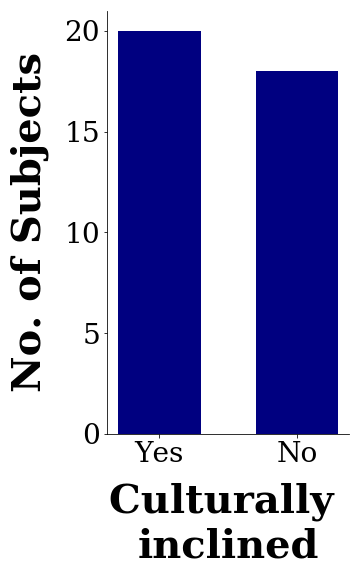}
    % \\$(vi)$
\end{minipage}
\hfill
\begin{minipage}{0.22\textwidth}
\centering
    \includegraphics[width=1.0\textwidth]{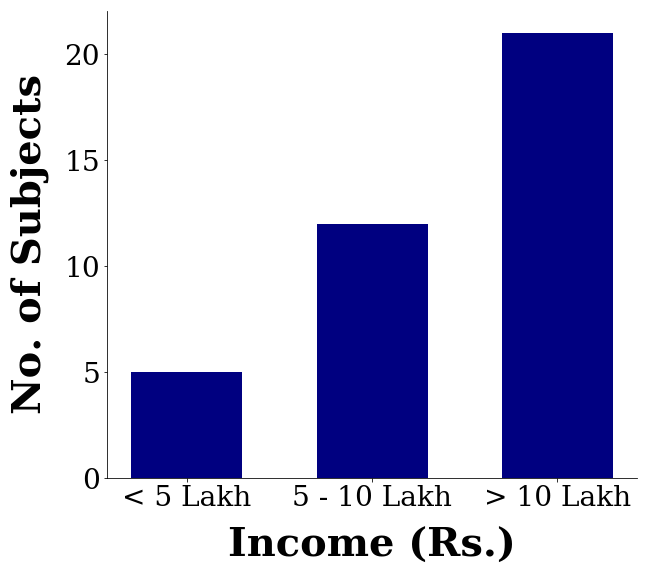}
    % \\$(ix)$
\end{minipage}
\hfill
\\
\hfill
\begin{minipage}{0.21\textwidth}
\centering
    \includegraphics[width=1.0\textwidth]{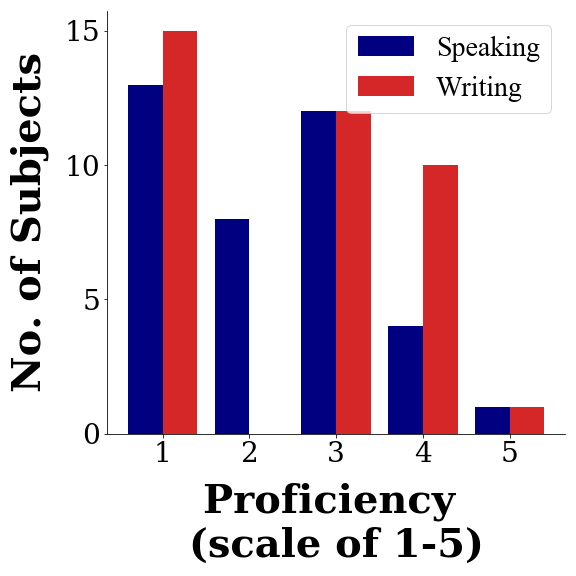}
    % \\$(vi)$
\end{minipage}
\hfill
\begin{minipage}{0.30\textwidth}
\centering
    \includegraphics[width=1.0\textwidth]{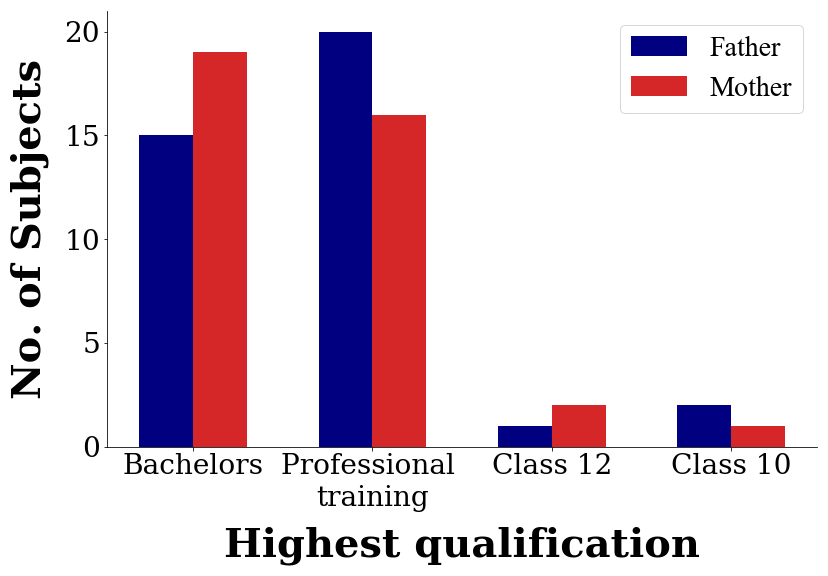}
    % \\$(x)$
\end{minipage}
\hfill
\begin{minipage}{0.47\textwidth}
\centering
    \includegraphics[width=0.850\textwidth]{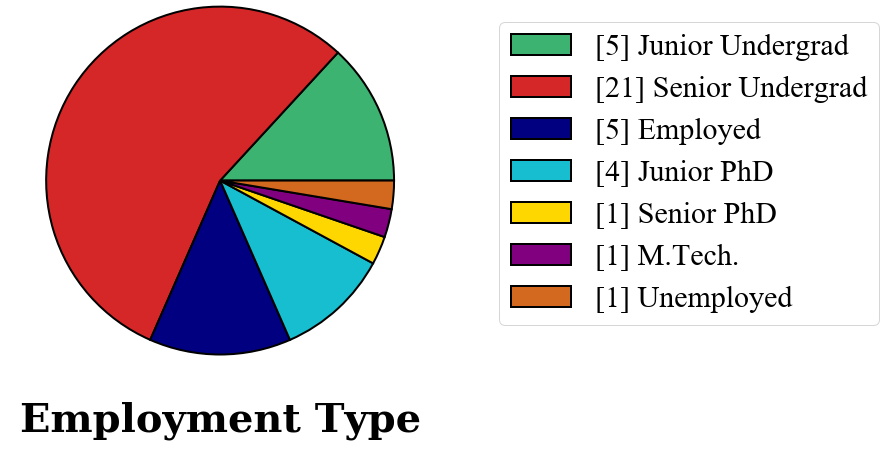}
    % \\$(xiii)$
\end{minipage}
\caption{Socio-Demographic Indicators: Distribution of the number of subjects belonging to different categories across these factors.}
\label{socio_demo_dist}
\end{figure*}
  
\vspace{1mm}    
 \noindent \textbf{Socio-demographic Information.} We collect socio-demographic indicators for all the participants involved in our study\footnote{Personally identifiable information (PII) has been anonymized to maintain the privacy of the participants.}. In order to do that, we designed a short questionnaire, shown in Table~\ref{socio-demo-ques}. Use of some of these indicators, like age and gender, was inspired by prior work \cite{goldberg1998demographic}, while several others, like if the subjects had a sibling and whether they lived with their family, have not been used previously. Our analysis reveals that these factors significantly impact the subjects' personality traits (see Section~\ref{socio-results}). A comprehensive view for the distributions of these socio-demographic variables is shown in Figure~\ref{socio_demo_dist}.

\begin{figure*}[!t]
\centering
\begin{minipage}{0.19\textwidth}
\centering
    \includegraphics[width=\textwidth]{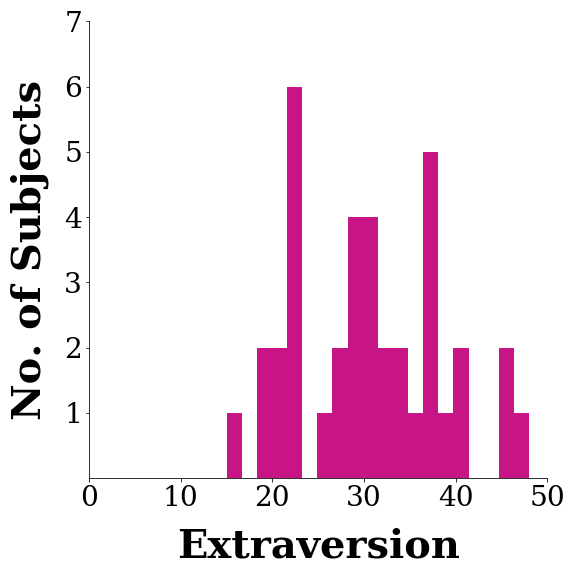}
    \\$(i)$
\end{minipage}
\hfill
\begin{minipage}{0.19\textwidth}
\centering
    \includegraphics[width=\textwidth]{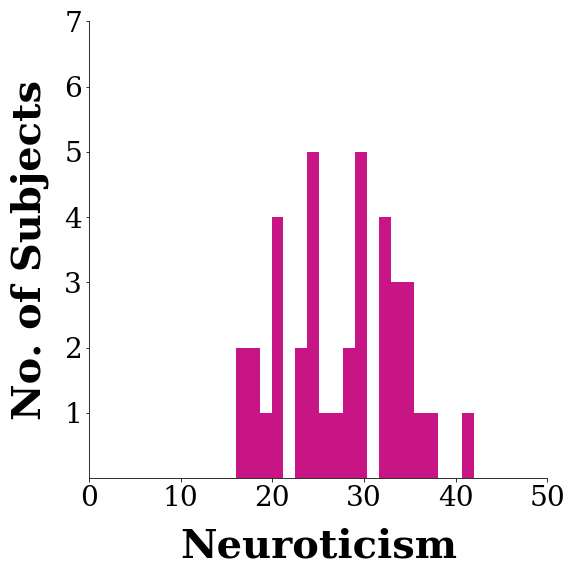}
    \\$(ii)$
\end{minipage}
\hfill
\begin{minipage}{0.19\textwidth}
\centering
    \includegraphics[width=\textwidth]{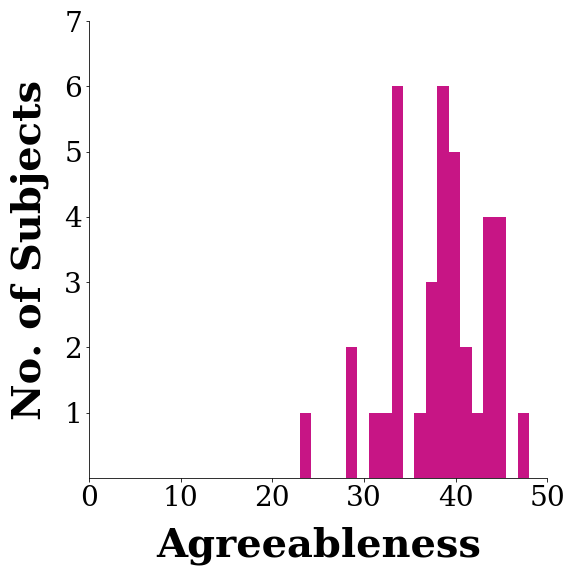}
    \\$(iiii)$
\end{minipage}
\hfill
\begin{minipage}{0.19\textwidth}
\centering
    \includegraphics[width=\textwidth]{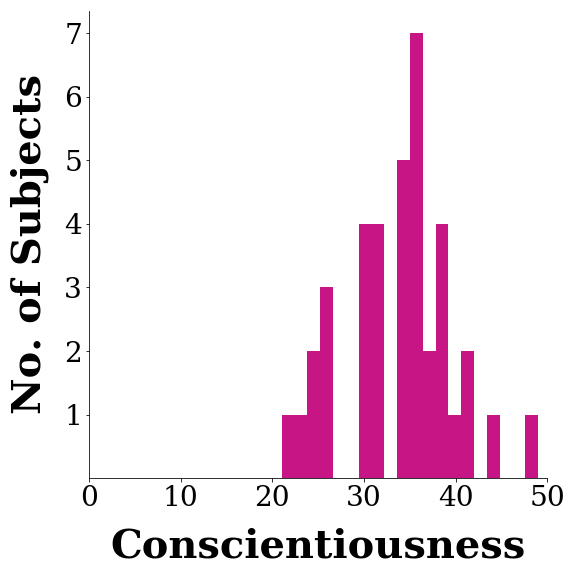}
    \\$(iv)$
\end{minipage}
\hfill
\begin{minipage}{0.19\textwidth}
\centering
    \includegraphics[width=\textwidth]{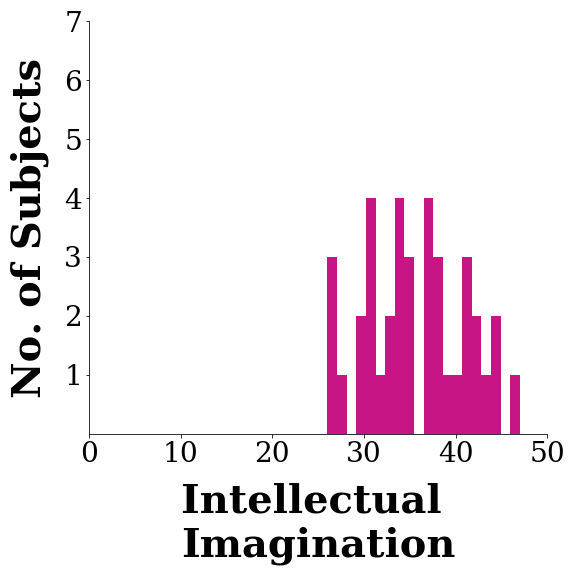}
    \\$(v)$
\end{minipage} 
\caption{Distribution of the raw Big Five personality trait scores.}
\label{raw_pers_scores}
\end{figure*}

\vspace{2mm}
 \noindent \textbf{Personality Traits.} Finally, we employ the widely used questionnaire designed by Goldberg \cite{goldberg1992development} to gather the Five-Factor personality traits of the participants. Figure~\ref{raw_pers_scores} shows the distribution of these scores for the subjects involved in the study.

\vspace{2mm}
\begin{table}[!t]
    \centering
    \resizebox{\columnwidth}{!}{
    \begin{tabular}{m{0.20\columnwidth}|m{0.40\columnwidth}|m{0.40\columnwidth}}
    \Xhline{1pt}

\textbf{Lexical Feature} & \textbf{Description} & \textbf{Examples}\\\hline
\textbf{Filled Pauses}  & Sounds made by the speaker to fill in the gaps while speaking.   & \textit{toh} (so), \textit{aise} (in this way), \textit{hmm}  \\\hline
\textbf{Discourse Markers} & Used to organize the dialogue into segments. & \textit{isliye} (because), \textit{to} (so), \textit{magar} (but), {matlab} (it means) \\\hline
\textbf{Suffix Words} & Words that the speaker elongates towards their end.  &\textit{areeyy haaan} (ohh yes!),  \textit{nahiii} (no), \textit{kyaaa yaar} (what is this)  \\\hline
\textbf{Repetition of words} & Speakers often do this during impromptu conversations. & \textit{jaise... jaise ki} (for example), \textit{magar}... \textit{magar ye to} (but this is)\\\hline
\Xhline{1pt}
    \end{tabular}}
    \caption{Lexical Annotations: Examples from the Hinglish transcriptions, along with their closest meaning in English.}
    \label{linguistic_annotations}
\end{table}

\section{Data Analysis}
\label{data_analysis}

\subsection{Analysis of Socio-Demographic Features}
\label{socio-results}

Socio-demographic factors like age, level of education, gender, and ethnicity have been known to impact an individual's personality \cite{goldberg1998demographic}. In this work, we statistically analyze several other indicators (Section~\ref{dataset}) that have not been considered in the past studies, and can potentially impact personality traits. Specifically, for each pair of a socio-demographic factor and one of the five personality traits, we use a two-sample Kolmogorov-Smirnov (K-S) test under the null hypothesis that the distribution of the scores of that trait conditioned on the indicator's values is comparable. Essentially, the test analyzes if there exists a significant difference in the Empirical Cumulative Distribution Functions (ECDF) of a random variable across the two samples under consideration. For instance, an individual may or may not speak in English when at home. Therefore, a two-sample K-S test evaluates the null hypothesis that the distribution of the personality scores for, say \textit{Extraversion}, when the person uses English at home, does not vary significantly from the distribution of the scores when he does not use English at home. For this particular example, however, we shall see shortly that the test rejects the null hypothesis. 

Since the two-sample K-S test compares only two distributions, we had to restrict the number of unique values of all socio-demographic factors to two, and handle the indicators that had more than two values accordingly. For features like \texttt{age}, \texttt{work experience},  \texttt{social media usage}, \texttt{public speaking skills}, and \texttt{writing proficiency}, we used the median values of their respective distributions to convert the values to binary, \textit{i.e.}, a value greater than the median would be labelled as \texttt{High}, and \texttt{Low} otherwise. For other features, like \texttt{mother's highest qualification},  \texttt{father's highest qualification}, \texttt{employment status}, and  \texttt{income}, we used dummy variables. For instance, \texttt{income $<$} \rupee \texttt{$0.5$ million} (\textasciitilde \texttt{USD 6600}) became a separate variable, with a value as $1$ if the income was in the range specified, and $0$ otherwise. Separate tests were run for all five personality traits.

\begin{figure*}[!t]
\centering
\begin{minipage}{0.19\textwidth}
\centering
    \includegraphics[width=\textwidth]{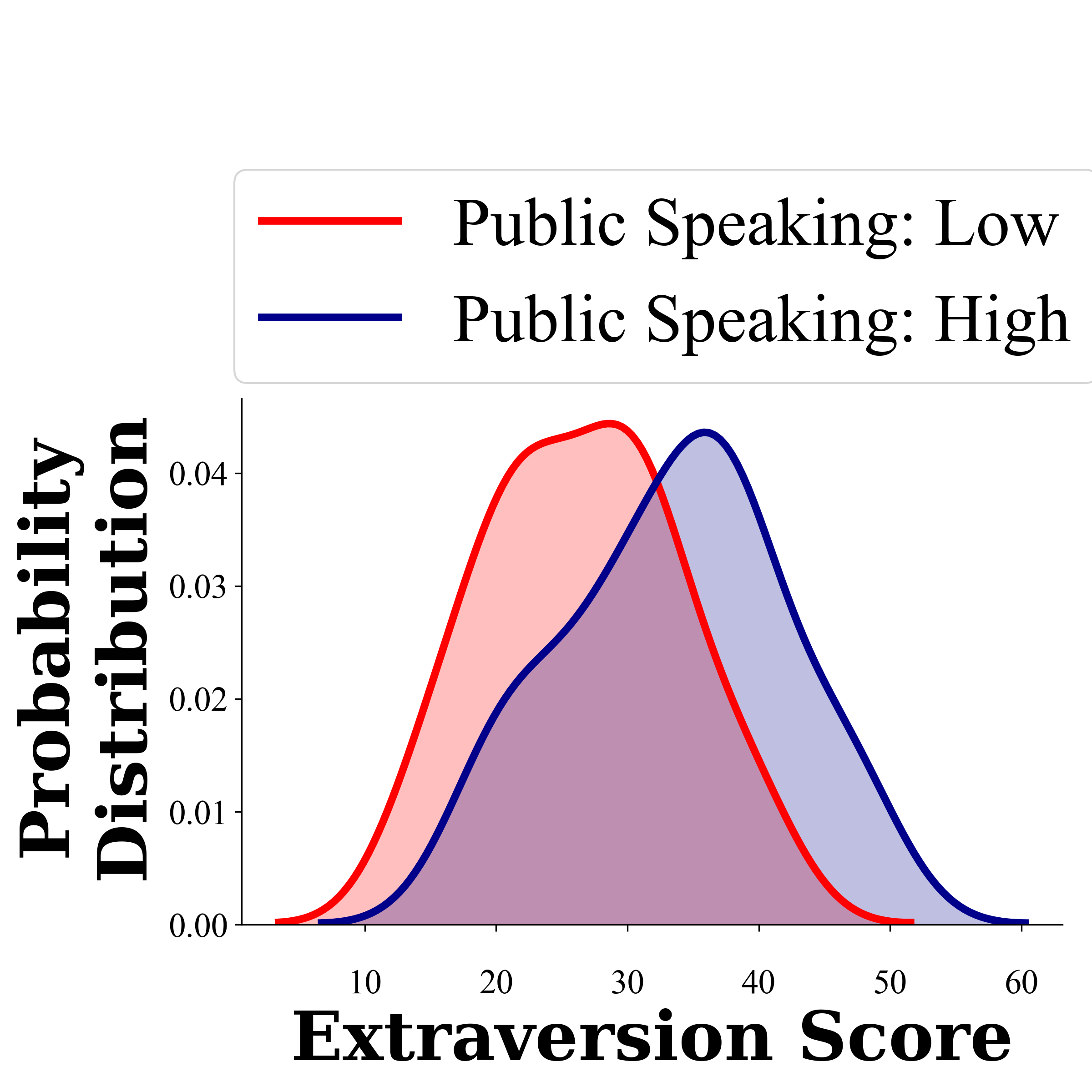}
    \\$(i)$
\end{minipage}
\hfill
\begin{minipage}{0.19\textwidth}
\centering
    \includegraphics[width=\textwidth]{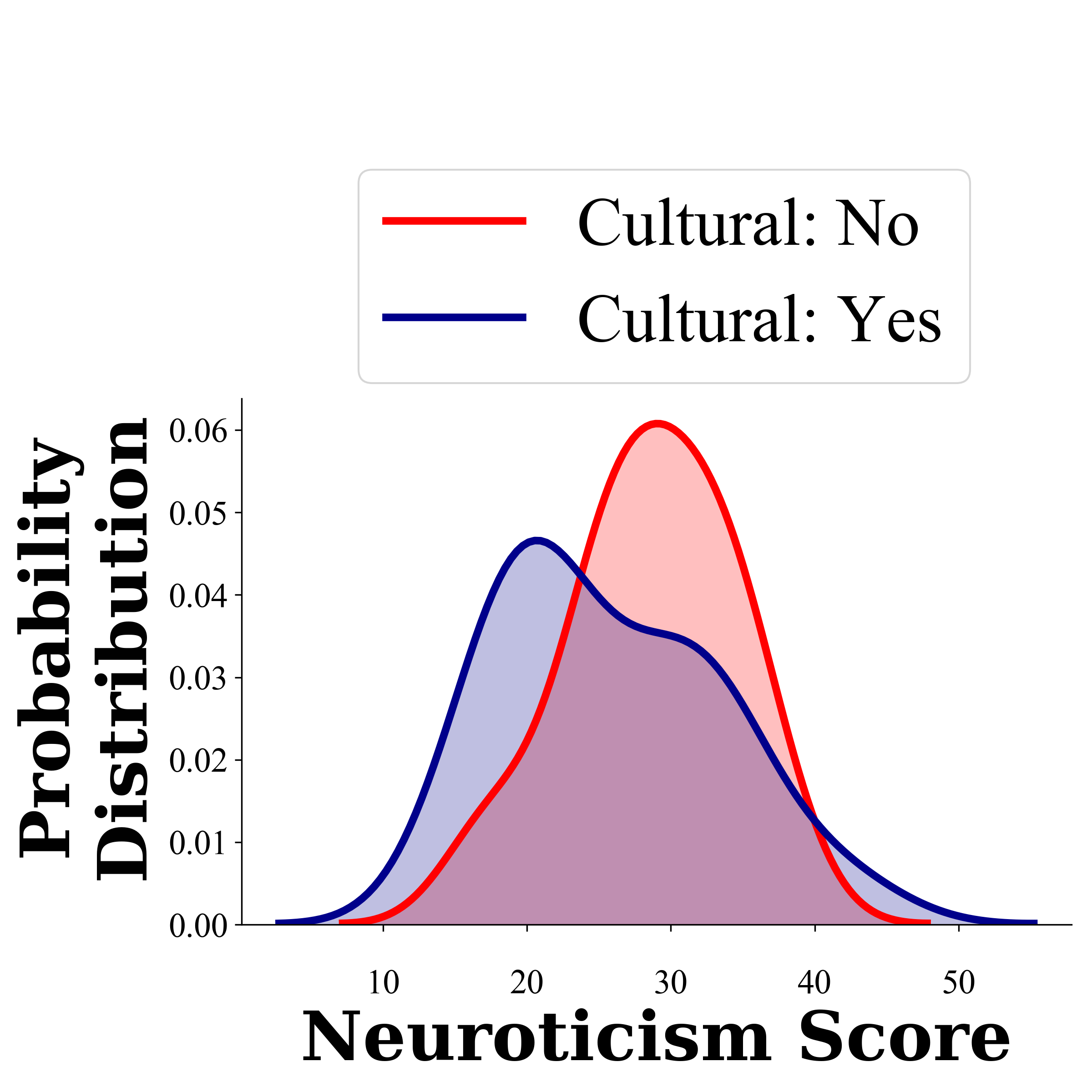}
    \\$(ii)$
\end{minipage}
\hfill
\begin{minipage}{0.19\textwidth}
\centering
    \includegraphics[width=\textwidth]{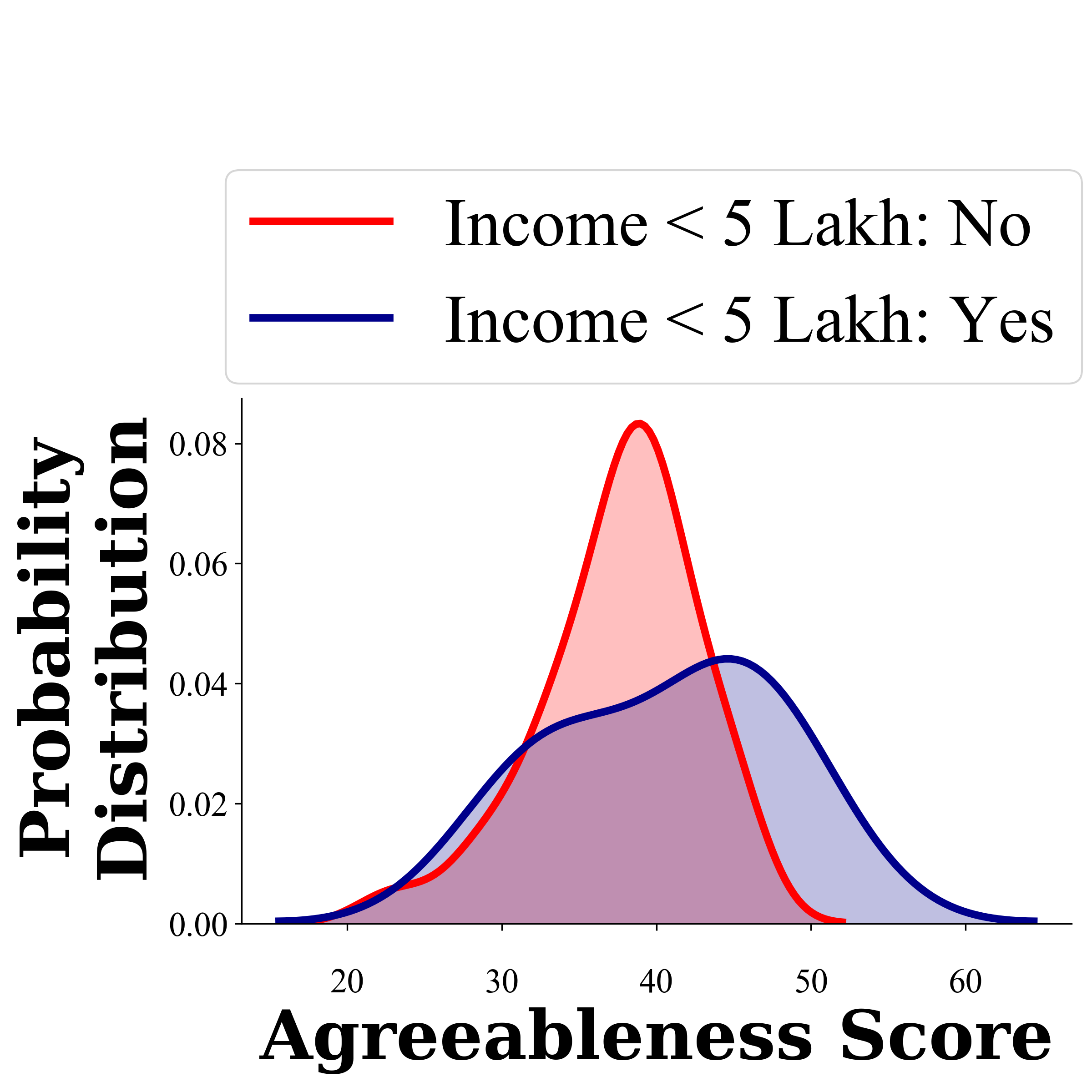}
    \\$(iiii)$
\end{minipage}
\hfill
\begin{minipage}{0.19\textwidth}
\centering
    \includegraphics[width=\textwidth]{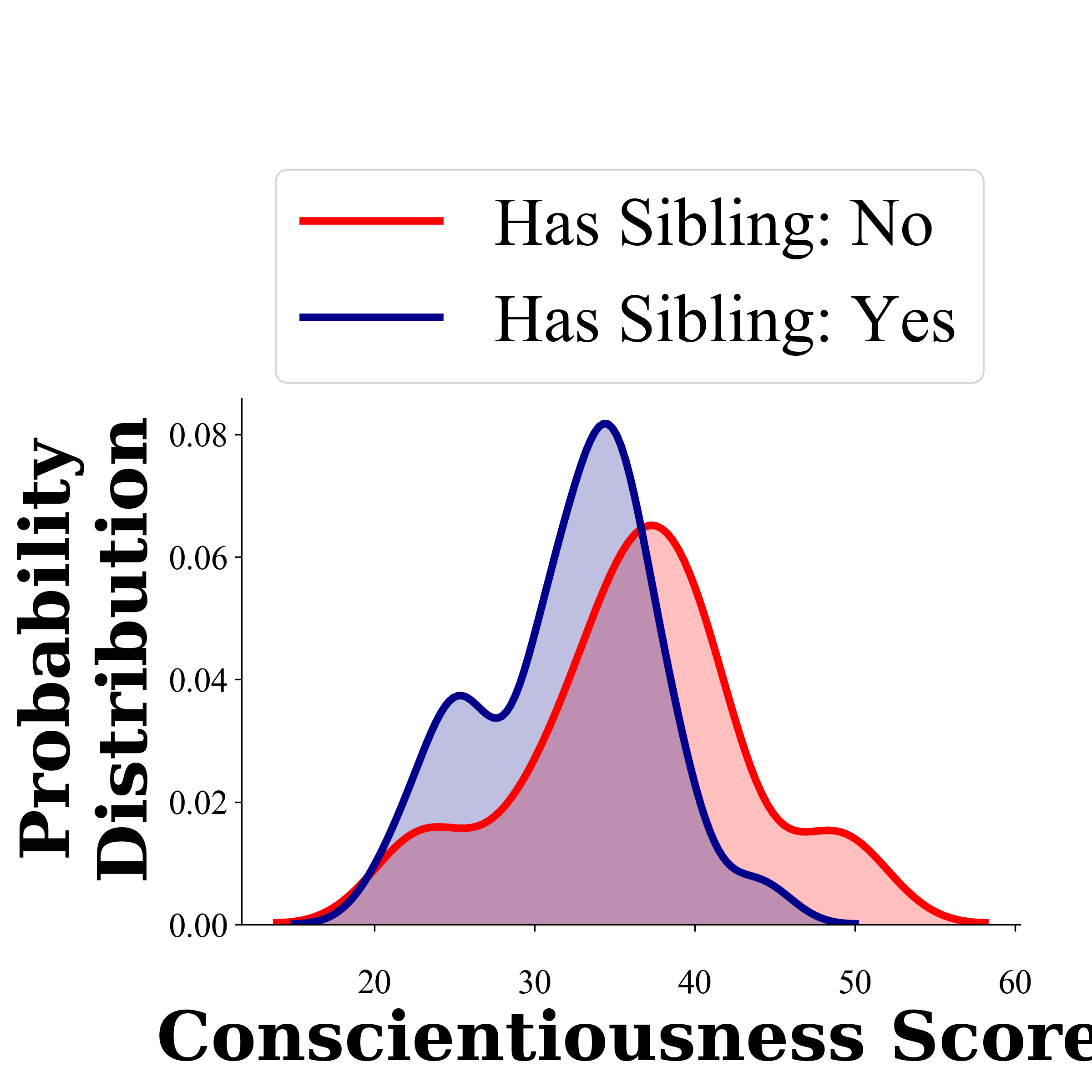}
    \\$(iv)$
\end{minipage}
\hfill
\begin{minipage}{0.19\textwidth}
\centering
    \includegraphics[width=\textwidth]{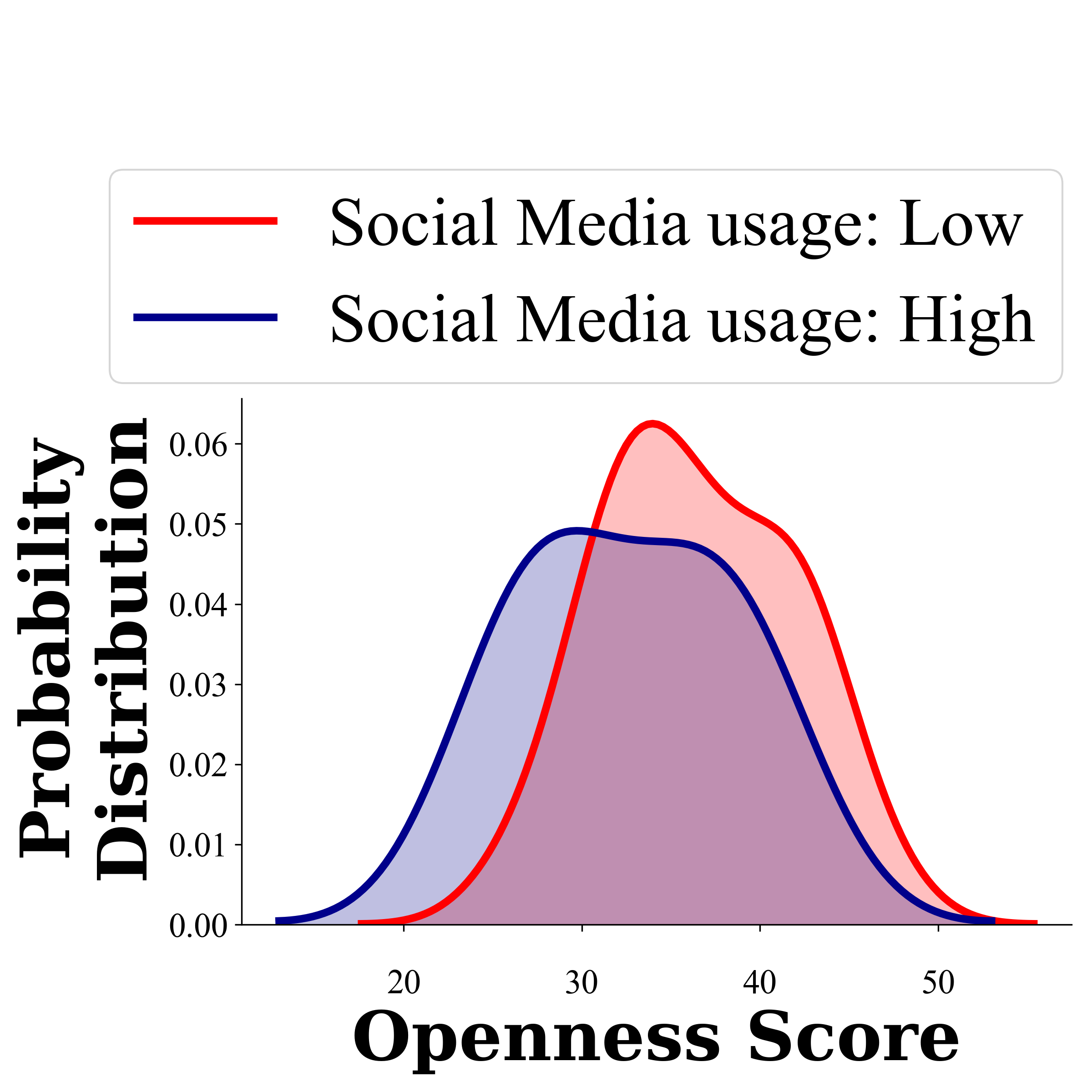}
    \\$(v)$
\end{minipage}
\caption{Probability Distribution Plots depicting the relation between different socio-demographic indicators and the Big Five personality traits: $(i)$ \texttt{EXT}, $(ii)$ \texttt{NEU}, $(iii)$ \texttt{AGR}, $(iv)$ \texttt{CON}, and $(v)$ \texttt{OPN}.}
\label{probability_dist}
\end{figure*}

Table~\ref{KS-socio-demo} presents the results of the K-S test conducted for all the Big Five personality traits separately. The features that were found as being of significant importance have been shown along with their corresponding value for the KS-Statistic ($\mathcal{K}$). In the interest of brevity, we have tabulated only $3$ significantly important indicators per trait. Furthermore, we also plotted the probability distribution functions of the indicators that we found particularly interesting. These plots are shown in Figure~\ref{probability_dist}. It can be seen that the subjects, who had high proficiency in public speaking, were most likely extroverts, as indicated by the distribution of the \textbf{\texttt{EXT}} scores. Furthermore, those who were more culturally inclined tended to be more stable emotionally, indicated by lower scores for \textbf{\texttt{NEU}}. Participants with annual income less than \rupee $0.5$ million (\textasciitilde USD $6600$) tended to be more straight-forward, as depicted by the \textbf{\texttt{AGR}} scores. Surprisingly, the \textbf{\texttt{CON}} score distribution revealed that subjects who did not have siblings were more disciplined than those who had siblings. Finally, based on the \textbf{\texttt{OPN}} distribution, those who had low social media usage were found to be more innovative.

\begin{table}[!t]
    \centering
    \resizebox{\columnwidth}{!}{
    \begin{tabular}{c|c|c}
    \Xhline{1pt}
    \textbf{Trait} & \textbf{Feature} & \textbf{KS-Statistic}\\\hline
    
    \multirow{3}{*}{\textbf{\texttt{EXT}}} & Father has Professional-training	 & \cellcolor{vred!35}	0.35\\
                                            & Public Speaking Proficiency & \cellcolor{vred!33} 	0.39\\
                                            & Writing Proficiency
 & \cellcolor{vred!42} 	0.42\\\hline
    
    \multirow{3}{*}{\textbf{\texttt{NEU}}} &  Cultural &\cellcolor{vred!28}	0.36\\
                                            &  Living with parents & \cellcolor{vred!26}	0.32 \\
                                            &  Social media usage	& \cellcolor{vred!47}	0.47\\\hline
                                            
    \multirow{3}{*}{\textbf{\texttt{AGR}}} & Income $<$ \rupee 0.5 million	& \cellcolor{vred!53} 0.48\\
     & Junior Undergrad Student  & \cellcolor{vred!44}	0.45\\
     & Age & \cellcolor{vred!19} 0.29\\\hline
     
     \multirow{3}{*}{\textbf{\texttt{CON}}} & Social Media usage & \cellcolor{vred!55}	0.49 \\
     & Sibling & \cellcolor{vred!18} 0.28\\
      & Mother's has Professional-training & \cellcolor{vred!28}	0.35\\\hline

\multirow{3}{*}{\textbf{\texttt{OPN}}} & Social Media usage	& \cellcolor{vred!35}	0.41 \\
& Writing Proficiency	& \cellcolor{vred!35} 0.41\\
& Sibling	& \cellcolor{vred!46} 0.46\\
\Xhline{1pt}
    \end{tabular}}
    \caption{K-S Test: The distribution of personality trait scores, conditioned on the respective socio-demographic features, varied significantly at a  5\% significance levels.}
    \label{KS-socio-demo}
    \vspace{-6mm}
\end{table}

\subsection{Lexical Analysis}
\begin{figure}[h!]
    \centering
    \includegraphics[width=0.30\textwidth]{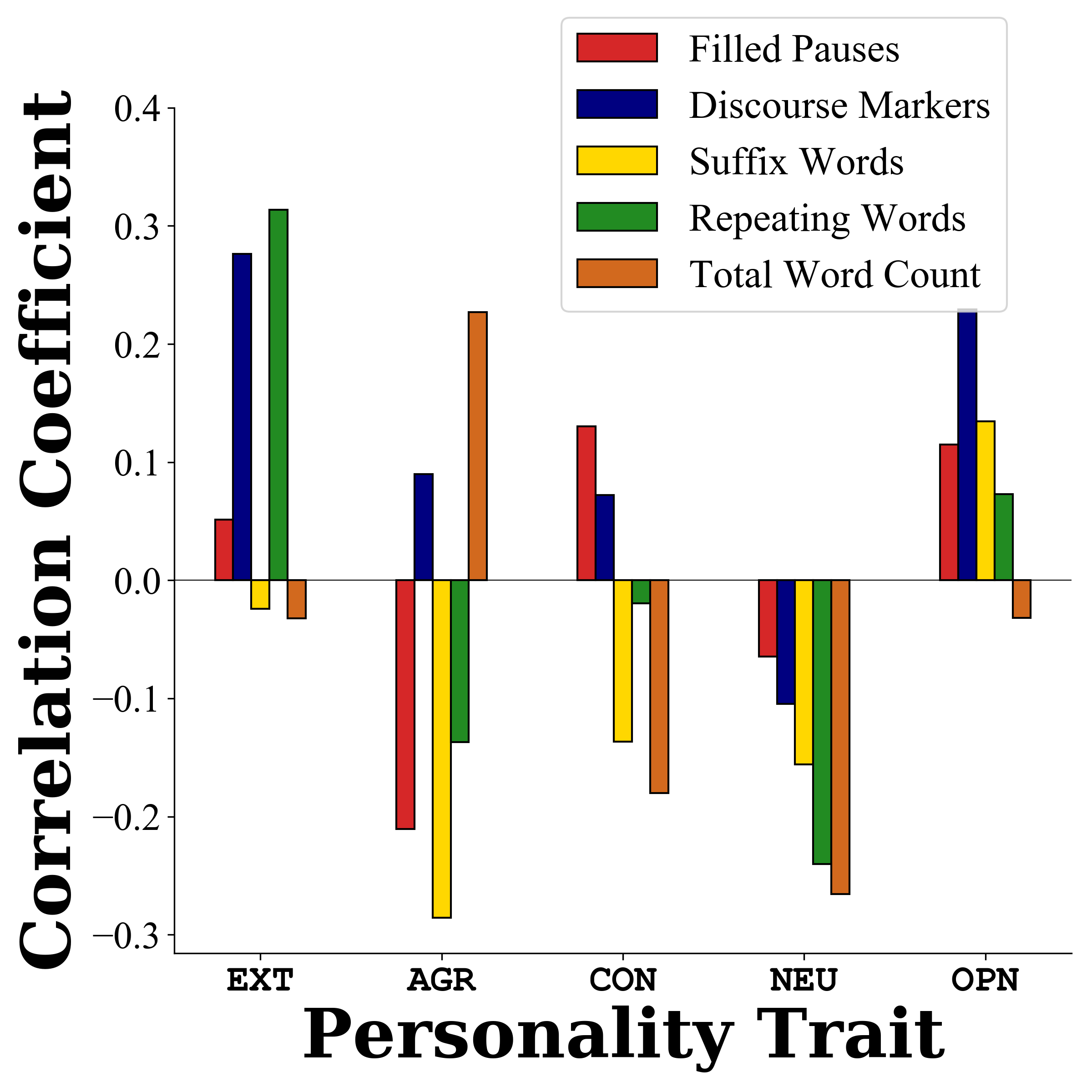}
    \caption{Correlation analysis between the Big-Five personality traits and frequency of filled pauses, discourse markers, suffix words, repeating words, and total words. }
    \label{lexical}
\end{figure}

In this section, we present the observations from our analysis of different kinds of lexemes (Table~\ref{linguistic_annotations}) used by the participants during their dialogues. Specifically, we perform a correlation analysis between the Big-Five personality trait scores and the frequency of filled pauses, discourse markers, suffix words, repeating words, and total words used by the participants. The plot in Figure~\ref{lexical} shows the Pearson correlation coefficients between the traits and the frequency of these words, depicting that the overall magnitudes for the correlation coefficients were not significantly high. Following are some other interesting observations we made:

\begin{itemize}
    \item There is a relatively high correlation between the use of discourse markers and the traits \textbf{\texttt{EXT}} and \textbf{\texttt{OPN}}. This also aligns with findings of previous studies \cite{doi:10.1080/10904018.2016.1202770} that established a positive relation between speaker backchannels and high extraversion scores.  
    
    \item Furthermore, frequency of repetition is positively correlated with \textbf{\texttt{EXT}}, and negatively with \textbf{\texttt{NEU}}. Low scores for \textbf{\texttt{NEU}} indicate high-levels of confidence in the subjects. This is an interesting finding that can be qualitatively analyzed further to draw parallels between individuals' psychological behaviors. 
    
    \item Subjects who had low scores for \textbf{\texttt{AGR}}, \textit{i.e.}, were less co-operative, used relatively more number of suffix words. In addition, these subjects also appeared to make more use of filled pauses, indicating verbal feedback. 
    
    \item Finally, the total number of words spoken by the subjects in their respective dialogues had a positive relationship with their level of confidence, shown by the negative coefficient for \textbf{\texttt{NEU}}. The opposite is true for \textbf{\texttt{AGR}}. Participants who used more words tended to have high scores.
\end{itemize}

\section{Recommendations for Future Work}
\label{recc_fw}
% Personality Assessment
% Others-
% Backchannel Opportunity Prediction, Conversational Agents
% Turn prediction: when should the agent start speaking.
%  
Given the diversity of the \textit{Vyaktitv} dataset, this section outlines some directions for potential future research. 

\begin{itemize}
    \item \textbf{Personality Trait Prediction.} A large body of research has been focussed on personality assessment \cite{sun2018personality, mawalim10.1007/978-3-030-21902-4_27, AHMAD20171964}. The dataset includes multimodal features, including the video and audio signals, as well as the transcriptions, which can be used to model personality in a multimodal setting \cite{shah2017multimodal, Mana2007MultimodalCO}. Since most of the research done so far has been on English datasets, \textit{Vyaktitv} acts as a rich, culturally different dataset in a low resource language that can be used to advance personality assessment research, and tackle challenges peculiar to such low resource languages.  For instance, lack of state-of-the-art word embeddings and impact of features that change with population, i.e., progressing towards cross-cultural personality assessment.     
    
    \item \textbf{Modelling and Design of Conversational Agents.} Research on Embodied Conversational Agents (ECA) has long been a primary center of focus for both the AI and HCI communities. Peer-to-Peer conversations have been extensively used for modeling a wide range of tasks that relate to the design of ECAs. These include, but are not limited to: predicting listener backchannels \cite{confpaperMMM, park2017telling}, disengagement prediction \cite{confpaperMMM}, and speaking turn prediction \cite{turnprediction}. With datasets such as \textit{Vyaktitv}, the research in the field of ECA design can potentially progress towards culture-sensitive agents, which can sense and accordingly modify their behavior based on the subjects around.  
\end{itemize}

% and the predictive models used can be used with other media and models in the hopes of obtaining a more accurate predictor for the big five personality traits. Further areas where the dataset proposed in the model could prove to be useful are listed below:
% \begin{itemize}
% \item Backchannel Opportunity Prediction. Robots can produce conditional listener responses and therefore deeply engage individuals as storytellers. An analysis of the dataset proposed can reveal the cues that can be decoded to find backchannel timing which can then be used to train the robot. \cite{7989266}
% \item Embodied conversational agents. The media content proposed in the dataset can be used to train machines to interact with humans and carry out a face-to-face social conversation with an individual in real time. \cite{Bickmore2005}
% \item Turn prediction. Previous studies \cite{turnprediction} have shown that factors such as gaze behaviour and respiration contribute to predicting the next speaker in a multi-speaker setting. An analysis of this dataset could further this and provide additional features that can be used for predicting when a user will end their speech to ensure smooth turn-taking.
% \end{itemize}

\section{Conclusion}
\label{conclusion}
In this work, we proposed a novel dataset, \textit{Vyaktitv}, for analyzing the personality traits of individuals through peer-to-peer social interactions. It is a multimodal dataset, where the subjects participate in pairwise conversations, and the primary language used is Hindi. We also collected a rich set of socio-demographic indicators from all the participants to study their impact on personality. In particular, we observed that factors like \texttt{income}, \texttt{public speaking skills}, \texttt{whether they live with their parents}, \texttt{social media usage}, and several others, were significant in distinguishing the distribution of the personality traits. Observations from the lexical analysis of the Hinglish transcriptions also indicated some interesting findings as to how the use of different types of filler words correlated with personality traits. Towards the end, we highlighted the diversity and richness of the multimodal Hindi dataset proposed in this work by discussing potential research directions that can be pursued.  It is not limited to personality assessment, but can also be used for several other challenging tasks that aim at designing human-computer interfaces, like Backchannel Opportunity Prediction (BOP) \cite{park2017telling}, Listener Disengagement Prediction \cite{confpaperMMM}, and modeling conversational agents \cite{10.1007/978-1-84882-215-3_13}. 

\section*{Acknowledgements}
We would like to acknowledge all the subjects who took part in the experiment, as well as the annotators who assisted in the transcribing the conversations. Furthermore, Jainendra Shukla and Rajiv Ratn Shah are partly supported by the Infosys Center for AI and the Center of Design and New Media at IIIT-Delhi.

\bibliographystyle{IEEEtran}
\bibliography{bibliography.bib}

\end{document}